\documentclass[aps,prl,twocolumn,superscriptaddress,showpacs,amsmath,amssymb]{revtex4}

\usepackage{graphicx}

\begin{document}

\title{Controlled transport of matter waves in two-dimensional optical lattices}

\author{Jasur Abdullaev}

\affiliation{Nonlinear Physics Center and ARC Center of Excellence for Quantum-Atom Optics, Research School of Physics and Engineering, The Australian National University,
Canberra ACT 0200, Australia}

\author{Dario Poletti}

\affiliation{Center for Quantum Technologies, National University of Singapore, Singapore
117542, Republic of Singapore}

\author{Elena A. Ostrovskaya}

\affiliation{Nonlinear Physics Center and ARC Center of Excellence for Quantum-Atom Optics, Research School of Physics and Engineering, The Australian National University,
Canberra ACT 0200, Australia}

\author{Yuri S. Kivshar}

\affiliation{Nonlinear Physics Center and ARC Center of Excellence for Quantum-Atom Optics, Research School of Physics and Engineering, The Australian National University,
Canberra ACT 0200, Australia}

\date{\today}

\begin{abstract}
We propose a method for achieving dynamically controllable transport of highly mobile matter-wave solitons in a  driven two-dimensional optical lattice. Our numerical analysis based on the mean-field model and the theory based on the time-averaging approach, demonstrate that a fast time-periodic rocking of the two-dimensional optical lattice enables efficient stabilization  and manipulation of spatially localized matter wavepackets via induced reconfigurable mobility channels. 
\end{abstract}


\maketitle

Controlled manipulation of stable, spatially localized matter waves, similar to routing of optical pulses in photonic devices, is a very attractive goal from the viewpoint of emerging integrated technologies based on the use of ultracold atomic gases - Bose-Einstein condensates~\cite{chip}.  In recent years optical lattices have been suggested as an instrument for such a control. Long-distance transport of ultracold atoms in a one-dimensional Bessel optical lattice potential was demonstrated experimentally~\cite{transport_1d}. Theoretical studies of matter-wave solitons, loaded into a rapidly driven asymmetric one-dimensional optical lattice, have demonstrated that such an "optical ratchet" could enable their mass-dependent transport \cite{dario}. Transport of matter waves in a two- or three-dimensional trapping geometry is a more complex and challenging task, especially considering the intrinsic instability of the self-localizing condensate with the negative scattering length.

Nonlinearly localized atomic wavepackets - matter-wave solitons, created in a two- or three-dimensional trapping geometry, are predicted to be stabilized by static optical lattices \cite{2d_3d_ol}. Other methods of stabilization rely on the time-periodic management of the condensate properties or non-local interaction between atoms \cite{santos}. Stability of the matter waves in two-dimensional optical lattices and under different conditions of the time-periodic management (see, e.g., \cite{burlak}) has been extensively studied. The main challenge still remains: To suggest an efficient method for non-destructive, dynamically controlled transport of the stabilized atomic wavepackets.

The problem of transport of localized nonlinear excitations is acutely posed in several cross-disciplinary areas of physics (see, e.g., Ref.~\cite{malomed_book} and references therein) because periodic systems, in general, greatly inhibit the mobility of the localized states. In particular, mobility of discrete solitons is an important problem in nonlinear photonics and lattice dynamics~\cite{ref1,ref2}.  In periodic structures and waveguide arrays, the motion of spatially localized optical beams is affected substantially by the lattice periodicity, so that the effective Peierls-Nabarro potential~\cite{ref2} induced by the lattice prevents their free motion. It was shown that moving solitary waves in discrete lattices require special type of saturable nonlinearity~\cite{ref3}, and can move only at some limited velocities~\cite{ref4}. Mobility of discrete optical solitons in two-dimensional photonic lattices was also found for  the case of quadratic nonlinear response that leads to parametric coupling of the fundamental and second-harmonic fields and results in their mobility after an initial kick ~\cite{ref5}.  

In this Letter we propose and analyze the mechanism for controlled transport of two-dimensional matter-wave solitons, created in a Bose-Einstein condensate of atoms with a negative scattering length. The transport is realized by means of a rocking two-dimensional optical lattice \cite{rocking}, where the term "rocking" refers to time-periodic shaking of the lattice \cite{rocking_theory}. Our analysis, based on the mean-field model, demonstrates that the fast rocking of the lattice enables dynamical creation of reconfigurable mobility channels for matter waves. Similar channels were previously studied in the context of lattices that do not coincide in dimensionality with the wavepacket and guide a matter wave along the fixed free spatial direction \cite{malomed}. Here we show that the rocking optical lattices enable both the efficient stabilization and dynamical ``routing"  of nonlinearly localized matter wavepackets.

To describe the dynamics of solitons in a Bose-Einstein condensate loaded into a two-dimensional lattice, we consider a mean-field model of an ultracold atomic cloud in a pancake trapping geometry with an optical lattice potential aligned with the weak trapping directions. As long as the transverse
collective modes of the condensate in the direction perpendicular to the optical lattice are not excited, the system can be treated as
two-dimensional and described by the Gross-Pitaevskii equation:
\begin{equation}\label{GP2D}
i\frac {\partial \psi}  {\partial t}+\nabla^2_\perp
\psi+|\psi|^2\psi+ V_{\rm OL}({\bf r},t)\psi=0,
\end{equation}
where $\nabla^2_\perp\equiv \partial^2/\partial {\bf r}^2$, and ${\bf r}=(x,y)$.
This model is derived by assuming the units of energy, length, and frequency:  $E_L=\hbar^2k^2_L/(2m)$, $a_L=1/k_L$, and $\omega_L=E_L/\hbar$,
respectively, where $m$ is the atomic mass, and $k_L$ is the
wavevector of the optical lattice. The three-dimensional mean-field model is reduced to the two-dimensional equation by assuming that the wavefunction is separable: $\psi_{3D}({\bf r},z)=\psi_{2D}({\bf r})\phi_{1D}(z)$, where $\phi_{1D}(z)$ is the normalized ground state wavefunction of a one-dimensional
harmonic potential with the trapping frequency
$\omega_z$. With these assumptions, the
wavefunction $\psi$ in Eq.~(\ref{GP2D}) relates to $\psi_{2D}$ as
follows: $\psi=\psi_{2D}\sqrt{g_{2D}}$, where
$g_{2D}=4\sqrt{\pi}(a_s/a_0)(\omega_z/\omega_L)^{1/2}$ is the renormalized
coefficient that characterizes interaction of the
condensate atoms with the s-wave scattering length $a_s$. The number of atoms
is given by: ${\cal N}=N/g_{2D}$, where $N=\int
|\Psi|^2 dx$ is the norm of the dimensionless
wavefunction.

For the $^7Li$ atoms with $a_s=-0.21$ nm, trapped using a CO$_2$ laser with   $\lambda=10.62$ $\mu$m \cite{BEC_soliton}, the physical values of the characteristic scales are:  $a_0=845$ nm, $\omega_0=2\pi \times 2039$ Hz. Consequently, the rescaled interaction coefficient is $g_{2D}=1.03\times 10^{-3}$, and the typical soliton has the norm $N\approx 10$, which corresponds ${\cal N}\sim 10^4$.

\begin{figure}[htb]
\includegraphics[width=7.5cm, keepaspectratio]{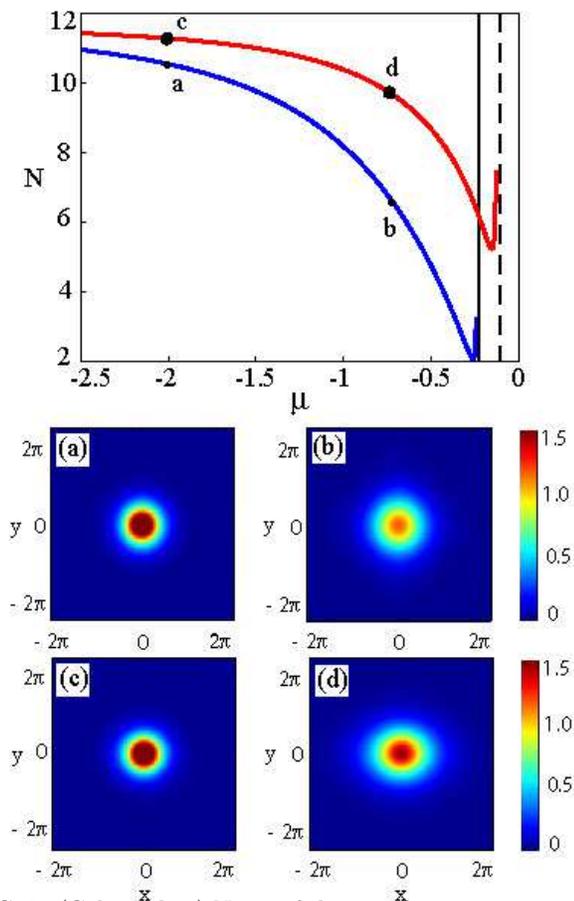}
\vspace{-0.6cm} \caption{(Color online) Norm of the wavefunction corresponding to a stationary soliton state in a rocking lattice potential $V_{\rm OL}({\bf r},t)$ at $t=0$ (curve $a-b$) and in a time-averaged potential $V_s({\bf R_s})$ (curve $c-d$) with $Y_0=0$, $X_0=2.4$ . (a-d) Spatial density structure of the matter-wave solitons at the marked points on the top panel.} \label{fig:families}
\end{figure}

The optical lattice potential in Eq. (\ref{GP2D}) is created by two pairs of counter-propagating laser beams and has the following form:
\begin{equation}\label{pot}
V_{\rm OL}({\bf r},t)=V_{x}\cos[x+X(t)]+V_y\cos[y+Y(t)].
\end{equation}
It is driven periodically, i.e. the phase detuning between the laser beams forming the lattice is modulated as: $X(t)=X_0\sin(\omega t)$, and $Y(t)=Y_0\sin(\omega t)$. By using Fourier decomposition, one can see that the optical lattice has both a static ``backbone" and a time-dependent component. This distinguishes transport in the potential (\ref{pot}) from the ratchet-type transport of matter-wave solitons \cite{dario}: Here the time-average force acting upon a wave-packet at any given point in space is non-zero.

\begin{figure}[floatfix]
\includegraphics[width=\columnwidth, keepaspectratio]{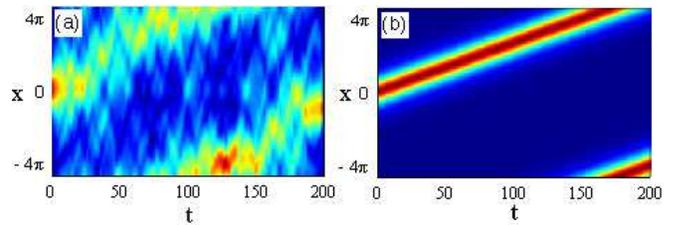}
\vspace{-0.6cm} \caption{(Color online) Dynamics of a moving soliton in the rocking lattice obtained by solving Eq. (\ref{GP2D}) numerically with periodic boundary conditions. Initial states are: (a) a stationary soliton of the potential $V_{\rm OL}({\bf r},0)$ at $\mu=-0.7$ [Fig. \ref{fig:families} (b)] and (b) a stationary soliton of the time-averaged potential $V_s({\bf R_s})$ at $Y_0=0$, $X_0=2.4$, $\mu=-0.7$ [Fig. \ref{fig:families}(d)]. } \label{fig:dynamics}
\end{figure}

At $\omega=0$ (or $t=0$), the static optical lattice $V_{\rm OL}({\bf r},0)$ supports stationary matter-wave solitons \cite{2D_soliton} in the form: $\psi({\bf r},t)=\Psi({\bf r})\exp(-\mu t)$ at the values of chemical potential, $\mu$, below the lower edge of the first spectral band (see Fig. \ref{fig:families}). These are found numerically by solving the stationary version of the model Eq. (\ref{GP2D}) \cite{yang}. According to the Vakhitov-Kolokolov stability criterion, such localized states are dynamically stable away from the band edge, where $d\mu/dN<0$ \cite{2d_3d_ol}. The mobility of the solitons is suppressed due to the energy difference between the nonlinear localized states with the same $N$ but different symmetry, that needs to be overcome in order to move the initially stationary state across the lattice. This is analogous to the Piers-Nabarro potential in discrete systems \cite{ref2}. On the other hand, it is the coupling to the lattice that both stabilizes the solitons against the collapse and allows us to manipulate the localized states by changing parameters of the lattice. The typical norm of a two-dimensional soliton considered in our dynamical simulations below is near the collapse threshold for a free soliton, which for our normalization is: $N^*=11.7$ \cite{burlak}.

To find the regime of enhanced mobility in the lattice we assume that a mobile soliton does not significantly change its shape and can be described solely by the dynamics of the center of mass, ${\bf r_0(t)}=(x_0(t),y_0(t))$. This assumption is valid provided that the driving frequency, $\omega$, is much greater than the eigenfrequency of soliton width oscillations ($\omega_0\sim 1$ in our system). Following the treatment in \cite{dario2}, the Hamiltonian theory that treats the soliton as a classical particle in an effective potential \cite{soliton_particle} can then be invoked to derive the equations of motion for the center of mass. The matter-wave soliton loaded into the lattice is approximated by the Gaussian function:
\[
\Psi_0({\bf r}(t),{\bf r_0}(t))=A\exp\left[-\frac{(x-x_0)^2+(y-y_0)^2}{2a^2}\right].
\]
The variational theory (see, e.g., \cite{burlak}) shows that this approximation is in good agreement with the exact stationary solutions for values of $\mu$ far from the band edge, where the soliton is well localized. In this case the soliton's amplitude and width are related as follows: $N=\pi A^2 a^2$, and $N=2\pi[2-V_0a^4\exp(-a^2/4)]$. The effective potential for the soliton-as-particle motion is then found as follows:
\[
V_{\rm eff}({\bf r_0},t)=\frac{1}{N}\int^\infty_{-\infty}|\Psi_0({\bf r},{\bf r_0})|^2V_{\rm OL}({\bf r},t)d{\bf r},
\]
where $d{\bf r}\equiv dxdy$, and the equations of motion for the center of mass, $d^2{\bf r_0}/dt^2=-dV_{\rm eff}({\bf r_0})/d{\bf r_0}$, read:
\[
\frac{d^2x_0}{dt^2}=V_xe^{-\frac{a^2}{4}}\sin(x_0-X), \,
\frac{d^2y_0}{dt^2}=V_ye^{-\frac{a^2}{4}}\sin(y_0-Y).
\]
The motion of the soliton's center of mass can be separated into the slow and fast components: $x_0(t)=X_s(t)+\xi(t)$, $y_0(t)=Y_s(t)+\eta(t)$, where the typical evolution of the slow variables ${\bf R_s}=(X_s,Y_s)$  occurs on the time scale $\tau\gg \omega^{-1}$. The time averaging procedure \cite{dario2,landafshitz} is then applied to derive equations of motion for the slow (compared to $\omega$) dynamics of a soliton center of mass in a driven lattice:
$d^2{\bf R_s}/dt^2=-dV_{\rm s}({\bf R_s})/d{\bf R_s}$. The time-averaged effective potential is separable: \begin{equation}
V_s({\bf R_{s}})=A_{sx}\cos(2X_s)+A_{sy}\cos(2Y_s),
\label{veff}
\end{equation}
with the strength explicitly depending on the driving frequency, $\omega$, and on the characteristic width of the soliton loaded into the lattice, $a$:
\begin{eqnarray}
A_{sx} &=& {\displaystyle \frac{V_x^2}{8\omega^2}e^{-\frac{a^2}{2}}\left\{4\left[J_1(X_0)\right]^2- \left[J_2(X_0)\right]^2\right\},} \nonumber \\
A_{sy} &=& {\displaystyle \frac{V_y^2}{8\omega^2}e^{-\frac{a^2}{2}}\left\{4\left[J_1(Y_0)\right]^2- \left[J_2(Y_0)\right]^2\right\},} \nonumber
\end{eqnarray}
where $J_n$ are Bessel functions of the first kind. We stress that the time-averaged potential is conservative even in the presence of driving at certain values of $(X_0,Y_0)$, namely such that $J_0(X_0)=0$ or $J_0(Y_0)=0$.

The time-averaged conservative potential, $V_s$, supports stationary matter-wave soliton solutions depicted by the family (c-d) in Fig. \ref{fig:families}. These solitons may be dynamically stable in the entire existence region of the solitons supported by the static lattice potential $V_{\rm OL}({\bf r},0)$ \cite{2D_soliton} due to the fact that the band edge in $V_s$ is shifted to lower values of the chemical potential, $\mu$ (dashed line Fig. \ref{fig:families}). This indicates that the driving may stabilize a moving solitons below the collapse threshold, as predicted in \cite{burlak} for amplitude-modulated optical lattices. In addition, for the same values of $\mu$, the mobility of the solitons supported by $V_s$ is expected to be enhanced compared to those in the static lattice $V_{\rm OL}({\bf r},0)$ due to the selectively suppressed spatial modulation.

 In order to confirm the predictions of improved mobility of the localized states in the driven lattice, we compare the dynamics of two initially stationary states corresponding to $\mu=-0.7$  and supported by $V_{\rm OL}({\bf r},0)$ and $V_s$ potentials (see Fig. \ref{fig:dynamics}). At the time $t=0$ both the initial states are given exactly the same initial velocity in the $x$-direction. Nevertheless, the soliton, supported by the static optical lattice $V_{\rm OL}({\bf r},0)$ (point $b$ in Fig. \ref{fig:families}) decays almost instantly, whereas the soliton supported by the $V_s$ in the driven lattice (point $d$ in Fig. \ref{fig:families})  moves and retains its shape [Fig. \ref{fig:dynamics} (b)].

\begin{figure}[htb]
\includegraphics[width=\columnwidth, keepaspectratio]{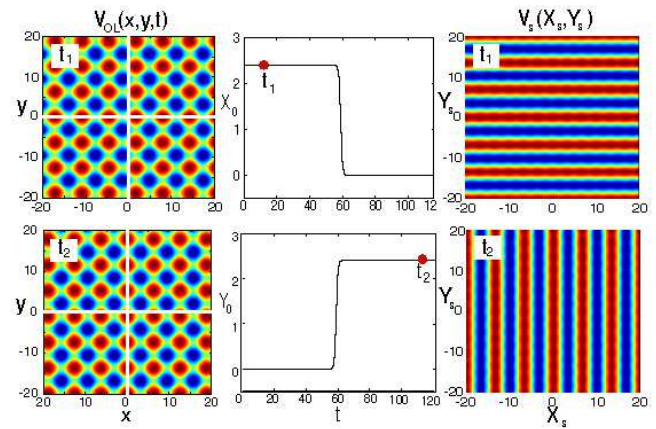}
\vspace{-0.6cm} \caption{(Color online) Dynamical control of the optical lattice parameters enabling creation and switching of the soliton mobility channels in the time-average potential. Left column: Rocking optical lattice potential, $V_{\rm OL}({\bf r},t)$ at the instances of time $t_1$ and $t_2$; the lines marking $x=0$ and $y=0$ are a guide to the eye. Middle column: Time dependence of the rocking amplitudes. Right column: Time-averaged lattice potential, $V_s({\bf R_s})$, corresponding to $Y_0=0$, $A_{sy}=0.5$, $A_{sx}=0.002$ (at $t_1$) and $X_0=0$, $A_{sx}=0.002$, $A_{sy}=0.5$  (at $t_2$).}
\label{fig:switching}
\end{figure}

\begin{figure}[htb]
\includegraphics[width=\columnwidth, keepaspectratio]{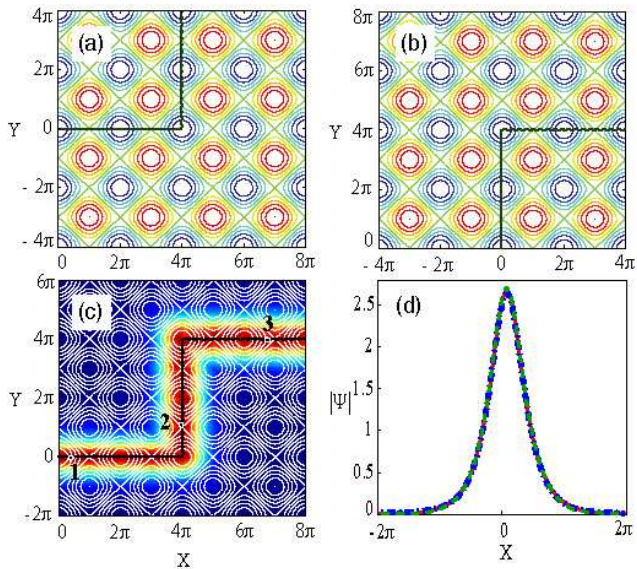}
\vspace{-0.6cm} \caption{(Color online) (a,b) Center of mass trajectory of a moving soliton corresponding to $\mu=-2.0$ in a rocking lattice potential $V_{\rm OL}({\bf r},t)$, superimposed onto the contour plot of the potential at $t=0$, and, in (c), onto the density profile of the moving soliton. Sharp turning points correspond to the switching of the mobility channels as shown in Fig. \ref{fig:switching}. (d) The profiles of the moving soliton corresponding to the points $1$ (solid), $2$ (dashed), and $3$ (dash-dotted) in (c).  }
\label{fig:trajectories}
\end{figure}

Next, we explain the basic principles of dynamical transport control of the mobile solitons in the driven lattice. While the lattice potential, $V_{\rm OL}({\bf r},t)$ is not static, and the positions of its maxima and minima are different at various times (see Fig. \ref{fig:switching}, left column), its action on the soliton center of mass during the periods of time $\tau \gg \omega^{-1}$ can be described by the conservative potential $V_s$. By examining the form of the time-averaged potential (\ref{veff}), one can see that the amplitude of the spatial modulation can be effectively suppressed in one of the orthogonal directions by controlling the amplitude of the periodic driving. The most dramatic example of such suppression is presented in Fig. \ref{fig:switching}. During the periods of time when the amplitudes of driving $X_0$ and $Y_0$ are kept constant (segments marked by the points $t_1$ and $t_2$ in Fig. \ref{fig:switching}, middle column), one of the amplitudes, $X_0$ or $Y_0$, is set to a value such that the time-averaged potential has a form of a two-dimensional lattice with the spatial modulation amplitude $A_{sx}$ or $A_{sy}$ radically suppressed, but not equal to zero (see Fig. \ref{fig:switching}, right column). Thus the mobility channels for a localized state are effectively created in the two-dimensional lattice along one of the orthogonal directions. This effective lattice is fully reconfigurable in real time, since direction of the mobility channels can be switched as shown in Fig. \ref{fig:switching}. During the switching time, which is short compared to $\omega^{-1}$, the concept of time-averaged potential is not applicable.

To demonstrate the transport of a matter-wave soliton in the dynamically reconfigurable driven lattice, we prepare the initial state in the form of the exact stationary soliton supported by the potential $V_s$ [point (c) in Fig. \ref{fig:families}]. We then load it into the lattice (\ref{pot}) driven with the frequency $\omega=10$ and the amplitudes $X_0=2.4$, $Y_0=0$, and let it evolve. At the time $t\approx 60$ the driving amplitudes are rapidly switched to the values $X_0=0$, $Y_0=2.4$, as shown in Fig. \ref{fig:switching} (middle panel). The resulting trajectory of the soliton's center of mass is shown in Fig. \ref{fig:trajectories}(a) against  the contours of the potential $V_{\rm OL}$ at $t=0$. The trajectory is easily reconfigurable by changing the sequence of switching [see Fig. \ref{fig:trajectories}(b)]. A more complex routing of the soliton along orthogonal directions in the lattice can also be realized, as shown in Fig. \ref{fig:trajectories}(c). In agreement with our initial assumption, the center-of-mass trajectory consists of slow (compared to $\omega$) motion perturbed by small amplitude fast oscillations. By examining the cross-sections of the soliton profile, we also observe that the localized state is only weakly affected by these dynamical manipulations [see Fig.\ref{fig:trajectories}(d)], which justifies our soliton-as-particle approach.

In conclusion, we have demonstrated, both analytically and numerically, that time-modulated optical lattices enable realization of clean, controlled transport of matter waves in more than one spatial dimensions. In doing so, we have used physical parameters relevant to experiments with the ultracold atoms. This work opens up a fascinating possibility to realize transport and routing of matter-wavepackets in lattice potentials of more complex symmetry, as well as in fully three-dimensional trapping geometry. Additional challenge is to achieve the transport of gap solitons supported by a lattice in a condensate with a positive scattering length, that are strongly "pinned" by the lattice and typically immobile. The work along these directions is currently underway.

This work was partially supported by the Australian Research Council (ARC). The authors are grateful to Dr. T. Alexander for stimulating discussions.

\end{document}